\documentclass[%
 reprint,
 amsmath,amssymb,
 aps,
pra,
nofootinbib,
noeprint
]{revtex4-2}

\usepackage{comment}
\usepackage{graphicx}
\usepackage{dcolumn}
\usepackage{bm}

\begin{document}

\title{Simulation of Cryogenic Buffer Gas Beams}

\author{Yuiki Takahashi}
\email{yuiki@caltech.edu}
\author{David Shlivko}
\email{Present address: Department of Physics,  Princeton University, Princeton, New Jersey 08544, USA}
\author{Gabriel Woolls}
\author{Nicholas R. Hutzler}
 \affiliation{Division of Physics, Mathematics, and Astronomy, California Institute of Technology, Pasadena, California 91125, USA}

\date{\today}

\begin{abstract}
The cryogenic buffer gas beam (CBGB) is an important tool in the study of cold and ultracold molecules.  While there are known techniques to enhance desired beam properties, such as high flux, low velocity, or reduced divergence, they have generally not undergone detailed numerical optimization.  Numerical simulation of buffer gas beams is challenging, as the relevant dynamics occur in regions where the density varies by orders of magnitude, rendering standard numerical methods unreliable or intractable.  
Here, we present a hybrid approach to simulating CBGBs that combines gas dynamics methods with particle tracing. The simulations capture important properties such as velocities and divergence across an assortment of designs, including two-stage slowing cells and de Laval nozzles. This approach should therefore be a useful tool for optimizing CBGB designs across a wide range of applications.



\end{abstract}

\maketitle

\section{Introduction}
There is considerable interest in physics with cold molecules because of their applications in a wide variety of areas, including quantum computing \cite{DeMille2002, Albert2020, Sawant2020}, quantum simulation of condensed matter systems \cite{Baranov2012,Bohn2017}, tests of fundamental laws of physics \cite{Safronova2018,DeMille2017,Chupp2019,Cairncross2019,Hutzler2020Review}, and ultracold and controlled chemistry  \cite{Krems2008,Ospelkaus2010Chemical,Balakrishnan2016,Bohn2017,Hu2019}. However, their internal degrees of freedom add additional complexity to the cooling and trapping of molecules, as compared to atoms.  There has been tremendous success creating ultracold molecules by associating ultracold atoms \cite{Ni2008,Bohn2017}, but many applications benefit from species containing atoms which are challenging to laser cool. Methods to cool generic molecules directly are therefore of interest, with the two most relevant techniques being supersonic beams \cite{Scoles1988} and buffer gas cooling \cite{Campbell2009Review,Hutzler2012}. These methods rely on collisions with a cold, inert gas such as helium or neon and can therefore cool a variety of species, from atoms to large complex polyatomics \cite{Patterson2010PCCP}, down to temperatures of $\sim1$~K.  These methods, especially buffer gas cooling, are also useful as starting points for further cooling, such as laser \cite{Tarbutt2019,McCarron2018}, sympathetic~\cite{Lara2006,Tscherbul2011Sympathetic}, evaporative~\cite{Stuhl2012,Reens2017}, or optoelectrical~\cite{Zeppenfeld2012,Prehn2016} cooling.

The cryogenic buffer-gas beam (CBGB) has emerged as a particularly powerful method for producing cold, slow, and intense beams of a variety of molecular species \cite{Maxwell2005,Hutzler2012}; it has been used for a range of spectroscopy \cite{Messer1984,Spaun2016,Porterfield2019, Santamaria2016} and precision measurement experiments \cite{ACME2018} and as the starting point for molecular laser cooling experiments \cite{Tarbutt2019,McCarron2018}. There are a number of established techniques to tailor CBGB properties to a particular application, such as hydrodynamic extraction \cite{Patterson2007} to create high-flux beams useful for precision measurements \cite{Hutzler2011,ACME2018}, or two-stage ``slowing'' cells \cite{Patterson2007,Lu2011} to create low-velocity beams suited to further slowing \cite{Anderegg2017} and trapping \cite{Lu2014}.  These techniques often trade one feature for another; for example, two-stage cells benefit from reduced velocities but also have lower brightness, while hydrodynamically extracted beams have higher flux but are ``boosted'' to higher velocities.

It would be useful to be able to rely on computer simulation to optimize CBGB design for any given application.  Direct simulation of buffer gas beams, however, is a computationally challenging problem.  The difficulty lies in the density regimes of the buffer gas and molecular\footnote{Though these sources are also useful for atoms, we shall refer to the species of interest as a molecule for simplicity.} species. CBGBs typically operate in the ``intermediate'' flow regime, where the mean free path between buffer gas collisions is not in the range of either the molecular or fluid limits \cite{Hutzler2012}.  Additionally, the species of interest often has a density orders of magnitude smaller than that of the buffer gas, which can cause out-of-equilibrium dynamics under typical operating parameters.  Accurately capturing the behavior of CBGBs therefore requires not only a reliable model of intermediate-regime flows, but also of dynamics on the molecular level.

In this manuscript, we utilize a two-step approach that explicitly incorporates these considerations to model a variety of CBGB implementations.  First, we model the steady-state buffer gas flow using Direct Simulation Monte Carlo (DSMC) \cite{Bird1994}, which enables computationally tractable simulations of intermediate-regime flows. Second, we trace individual particles of the molecular species through the background buffer gas flow using a random collision model.  We find that this approach is able to reproduce a number of important and non-trivial features of a variety of CBGB designs \cite{Hutzler2012}, such as aerodynamic focusing and the effects of slowing cells and de Laval nozzles.  These results indicate that the two-step approach correctly models both the intermediate-density regime of the cell and the low-density regime of the beam itself, offering a robust tool for CBGB optimization across a wide range of applications.

Several previous works have modeled the behavior of molecules in a buffer gas cell using a variety of methods and their combinations, such as particle tracing with random walk processes \cite{Maxwell2005, Gantner2020}, continuum fluid simulation \cite{Singh2018,Truppe2018} in combination with diffusion \cite{Bulleid2013}, direct simulation Monte Carlo (DSMC) \cite{Schullian2015, Doppelbauer2017}, and Self-Consistent Mean Field DSMC (SCMFD) \cite{Schullian2019}. The last of these is most similar to the methods employed here, though all of these previous works share some similarities with our approach.

\section{Methods}

\begin{figure}[ht]
\begin{center}
\includegraphics[width=0.47\textwidth]{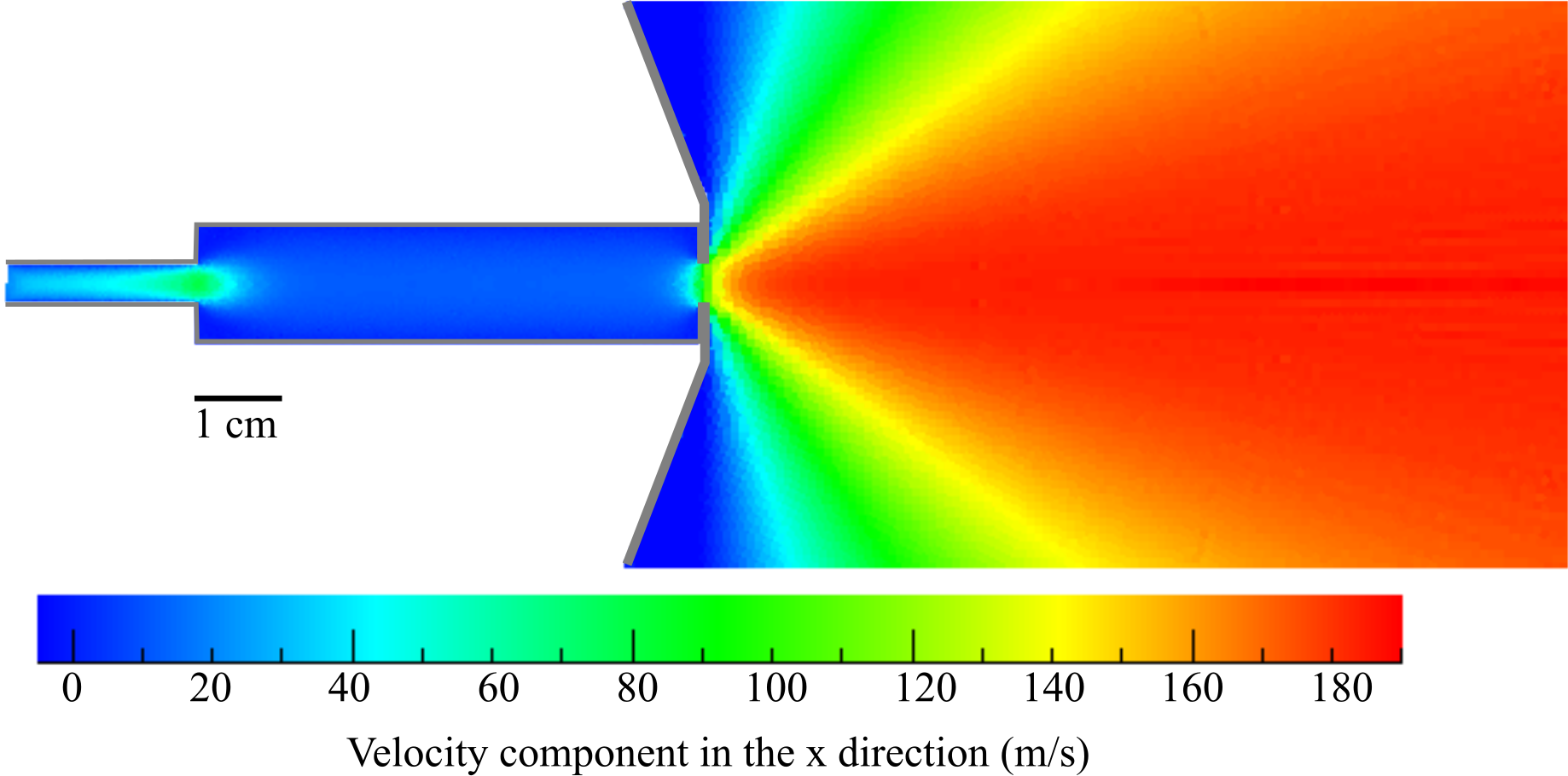}
\end{center}
\caption{Spatial variation of the forward velocity of the 4 K He buffer gas at around 20 SCCM in a single-stage cell computed with DS2V.  In this and all other cells, the buffer gas enters on the left and exits through a 5 mm diameter, 0.5 mm thick aperture before expanding into vacuum.  Gray lines indicate 4 K surfaces.}
\label{singleds2v}
\end{figure}

In this paper, we focus on simulations of a CBGB source consisting of a cryogenic volume or ``cell'' with cm-scale dimensions connected to a buffer gas inlet tube and an opposing exit aperture, as shown in Figure \ref{singleds2v}.  In steady-state, the balance between flows through the inlet and aperture creates a roughly stagnant density of buffer gas in the cell.  Hot atoms or molecules are introduced through ablation of a solid precursor or through a heated inlet, thermalizing with the cold buffer gas and exiting through the aperture as a molecular beam.  Details of typical CBGB sources, along with relationships for quantities such as flow, density, diffusion, \textit{etc.}, can be found in Ref. \cite{Hutzler2012}.

In this section, we present a two-step approach to simulating the dynamics of the CBGB. The first step is a DSMC simulation that computes the properties of the buffer gas alone, including its equilibrium flow velocities, temperatures, and densities throughout the cell and beyond the aperture. The second step is a particle-by-particle simulation of the molecular species as it traverses the cell while interacting with the steady-state buffer gas and emerges as a beam.

\subsection{Direct Simulation Monte Carlo}
The DSMC method is a statistical technique for simulating gas dynamics via the Boltzmann equation \cite{Bird1994}. 
DSMC is particularly useful for modeling the density regimes of typical cryogenic buffer gas beam experiments, where gases are too dilute for the Navier-Stokes Equations to be applicable, but too dense for a tractable computation of each atom’s trajectory. In the DSMC algorithm, each simulated particle represents a large number of identical particles, so that the overall distribution in phase space is maintained but fewer computations must be made to recover an accurate description of the flow dynamics. These particles are grouped into ``cells'' (not to be confused with the cryogenic cell containing the buffer gas and species) and nearby particles within a cell are allowed to collide and exchange momentum. Beyond these general features, the mechanics of the DSMC algorithm are largely adjustable to best fit the context of a specific system.

The DSMC software we utilize in this work is DS2V\footnote{\MakeLowercase{http://www.gab.com.au/index.html}}. We assume an axially symmetric flow to simulate a cylindrical buffer gas cell and allow the buffer gas to flow in through the inlet at rates between 1 and 130 SCCM (standard cubic centimeter per minute, or about $4.5\times10^{17}$ atoms per second). Typical number densities and velocities of this flow are on the order of 10$^{15}$ cm$^{-3}$ and 10 m/s. The simulations produce detailed information about the properties of the buffer gas, such as temperature, number density, and flow velocity, on a mesh of points within the specified geometry of the buffer gas cell and the area beyond its aperture. The walls of the cell in the simulation are kept at 4 K and are assumed to be diffuse reflectors for the buffer gas and perfect absorbers for the molecular species, since the latter has negligible vapor pressure at cryogenic temperatures. The cell is initially empty (vacuum), and the simulation runs until it converges to a steady state. The mass flow rate into the cell is explicitly calculated by integrating the flux over the surface of the inlet tube aperture. We model the helium-helium collisions as hard spheres with cross-section $\sigma_{b-b}= 2\times10^{-15} $cm$^{2}$  \cite{Kestin1984}. Though the true cross section depends on collision energy \cite{Sharipov2017}, we have varied the buffer gas collision cross-section by a factor of $\sim$5 in either direction and confirmed that the main features and trends are relatively unaffected. Future improvements of this technique could include more accurate collision models, but we find that the hard sphere approximation is sufficient for the present purpose.

\subsection{Particle Tracing}
Once the properties of the buffer gas have been determined by DSMC, we trace the paths of individual molecules as they move through the steady-state buffer gas flow. Because the density of the molecular species is significantly lower than that of the buffer gas in typical experiments, each molecule's path is traced independently of the others, and collisions occur only with the buffer gas. The general structure of the particle tracing algorithm is as follows:
\begin{enumerate}
	\item \emph{Specify the molecule's initial state.} The position \textbf{x} and velocity \textbf{v$_\text{s}$} of the molecular species $(s)$ are either specified explicitly or drawn randomly from a desired distribution. The initial conditions used to generate our results in Section \ref{results} will be stated and motivated therein.
	
	\item \emph{Compute the duration of the present time step.}
	We use a dynamic time step chosen\footnote{Taking smaller (more precise) time steps does not noticeably impact the results of these simulations, as long as the probability of a collision in a time step is $\lesssim 0.3$.}  such that the probability of collision at every step is $p=0.1$. 
	To accomplish this, we compute the collision rate of the molecular species,
    \begin{equation}
    	\Gamma = \sigma_{b-s} n \langle v_\text{rel} \rangle,
    \end{equation}
    where the hard-sphere elastic collision cross section between the buffer gas $(b)$ and the molecular species, $\sigma_{b-s}$, is set to a benchmark value of $3\times10^{-14}\text{cm}^{2}$ unless specified otherwise. The buffer gas number density $n = n(\textbf{x})$ is interpolated from DSMC data at the molecule's position, and the mean velocity of buffer gas atoms relative to the molecular species $\langle v_\text{rel} \rangle$ is computed from the molecule's velocity and the interpolated buffer gas temperature and flow velocity at its position using Eq. (\ref{vrel}). The desired time step is then $\Delta t=p/\Gamma=0.1/\Gamma.$
	
	\item \emph{Update the molecule position and velocity.}	We first update the molecule's position given its current velocity, $\textbf{x} \mapsto \textbf{x} + \textbf{v$_\text{s}$}\Delta t$.	Next, if a collision occurs (which is decided at random with probability $p=0.1$), we randomly choose a collision partner based on the local buffer gas properties and compute the resulting velocity change of the species.  This process is described in detail in the appendices, but we provide a brief summary here.
	
	First, we randomly sample the thermal velocity \textbf{u} of a colliding buffer gas atom from the conditional distribution
	\begin{equation}
	    f(\textbf{u}|\text{coll}) \propto v_\text{rel} \times f_\text{MB}(\textbf{u}),
	\end{equation}
	where 
	\begin{equation}
	    v_\text{rel} \equiv |\textbf{v$_\text{s}$} - \textbf{v$_\text{b}$}| = |\textbf{v$_\text{s}$} - (\textbf{u} + \textbf{v}_\text{flow})|
	\label{vflow_lab}
	\end{equation}
	is the relative velocity between the molecule and the randomly chosen buffer gas atom, $f_\text{MB}(\textbf{u})$ is the Maxwell-Boltzmann distribution, and $\textbf{v}_\text{flow}$ is the buffer gas flow velocity extracted from DSMC. This weighted distribution accounts for the greater likelihood of collisions involving atoms with higher relative velocities. The process by which we sample from this distribution is detailed in Appendix \ref{bayes_sec}. 
	
	Once a buffer gas atom velocity is determined, we employ a hard-sphere elastic collision model to compute the impulse felt by the molecule when colliding with this buffer gas atom with a random impact parameter. The process of sampling an impact parameter and computing the post-collision velocity of the molecule is detailed in Appendix \ref{boyd}.

	\item \emph{Repeat Steps 2-3 until the molecule reaches a wall or a specified finish line.} The simulation ends if the molecule hits a wall of the specified geometry (at which point it sticks to the surface) or reaches an end point of the simulation space. 
\end{enumerate}

\section{\label{results}Results}

The goal of this study is to model the qualitative behaviors of the CBGB in a number of different experimental configurations. We will begin with a single-stage cell fixed at typical experimental dimensions and add features incrementally, such as a second slowing stage and a de Laval nozzle. In this way, we can explore the direct effects of those changes on the beam. We will predominantly be simulating SrF molecules in a 4 K helium buffer gas~\cite{Barry2011} to model typical conditions of previous experiments, which range over both heavier and lighter species~\cite{Hutzler2012}.

\subsection{Single-stage cell}

The single-stage buffer gas cell is the standard design for producing cold molecules using the CBGB method \cite{Maxwell2005,Hutzler2012}. The flow of the buffer gas is usually operated in the ``hydrodynamic'' enhancement regime, where hydrodynamic forces push the molecules out of the cell and into the beam before they have time to diffuse and stick to a cell wall \cite{Patterson2007}.

Figure \ref{singleds2v} shows the forward velocity of the 4 K He buffer gas with around 20 SCCM input flow in the single-stage cell used in this simulation. The cell is cylindrical with a 12.7 mm internal diameter and 53 mm length. 4 K helium gas flows into the cell through an inlet tube whose internal diameter is 4 mm and length is 20 mm. The exit aperture has a 5 mm diameter and 0.5 mm thickness.

\subsubsection{Flow parameterization}

 \renewcommand{\Re}{\mathcal{R}}
 \newcommand{\Kn}{\mathcal{K}}
 \newcommand{\Ma}{\mathcal{M}}

\begin{figure}[ht]
\setlength{\belowcaptionskip}{-12pt}
\begin{center}
\includegraphics[width=0.47\textwidth]{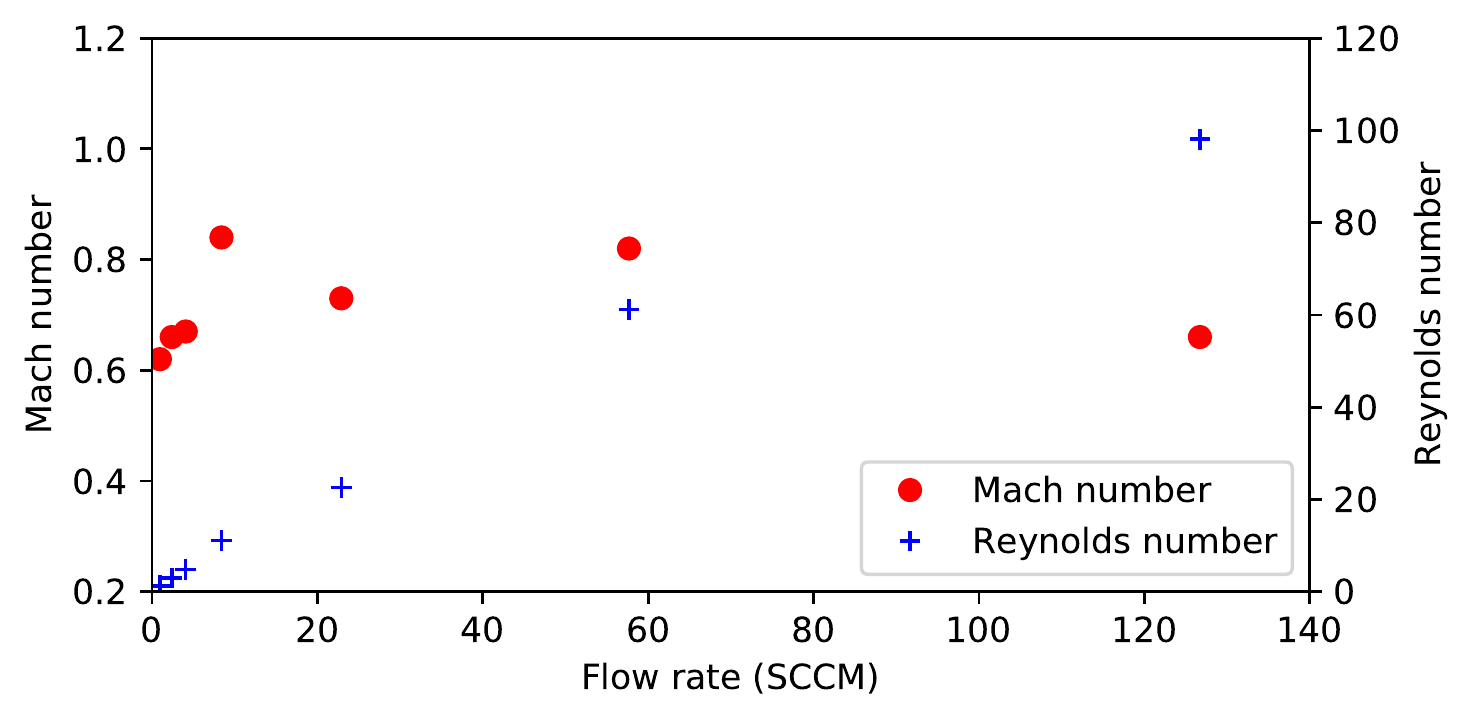}
\caption{Reynolds number $\Re$ and Mach number $\Ma$ at the aperture versus 4 K He flow (SCCM) for a single-stage cell.}
\label{MaRe}
\end{center}
\end{figure}

In discussing our results, it is useful to relate several quantities to describe the buffer gas flow \cite{Hutzler2012}, such as the mean free path between buffer gas collisions $\lambda$ and Knudsen number $\Kn=\lambda/d_{aperture}$.   Another common parameter is the Reynolds number $\Re$, which can be related \cite{Sone2007,Hutzler2012} to microscopic quantities via 
\begin{equation}
\label{MKReq}
    \Ma \approx \frac{1}{2}\Kn\Re,
\end{equation}
where the Mach number $\Ma$ is the ratio of the buffer gas flow velocity to the local speed of sound. $\Ma$ is typically estimated as being $\approx 1$ at the aperture for these sources, which our DSMC simulations show to be reasonably accurate (see Fig.  \ref{MaRe}). However, the Mach number of the buffer gas varies by around an order of magnitude within a few diameters of the exit aperture, as expected for a free jet expansion \cite{Pauly2000}.  The parameters $\Kn$ and $\Re$ are frequently used to classify flows into the molecular $(\Re\lesssim 1)$, intermediate $(\Re\approx 1-100)$, and hydrodynamic $(\Re\gtrsim 100)$ limits \cite{Hutzler2012}.

We can use eq. (\ref{MKReq}) to extract a linear relationship \cite{Hutzler2012}  between flow $f$ and Reynolds number $\Re$ with slope $\Re/(f/\mathrm{SCCM})\approx 0.8$, as shown in Fig. \ref{MaRe}.  Plots presented here are parameterized in terms of buffer gas flow, but we include this conversion as some CBGB papers use $\Re$ instead.

\subsubsection{Extraction}

The extraction fraction is defined as the total number of molecules emitted into the beam divided by the total number of molecules initially present in the cell. For hydrodynamically extracted beams, this is typically 10-50\% \cite{Patterson2007,Hutzler2012}. This efficient extraction, combined with the buffer gas density being too low for appreciable losses from reactions or clustering, is one of the reasons why CBGB sources are so bright, especially for refractory and radical species.  In the context of experiments where molecules are introduced into the cell through laser ablation, which is common for these types of challenging species, extraction will depend on complex ablation dynamics that set the initial spatial distribution of molecules \cite{Skoff2011PRA, Tarallo2016}. While our computational methods could be employed to investigate the post-ablation thermalization itself,\footnote{Our simulations of the thermalization process indeed recovered the time constant estimated in \cite{Hutzler2012}.} we choose in this paper to focus on exploring the molecular dynamics post-thermalization.  Under typical conditions, the thermalization process occurs rapidly and yields a fairly uniform distribution of molecules throughout the cell \cite{Skoff2011PRA}, making this a valid starting point for simulation.

\begin{figure}[ht]
\includegraphics[width=0.47\textwidth]{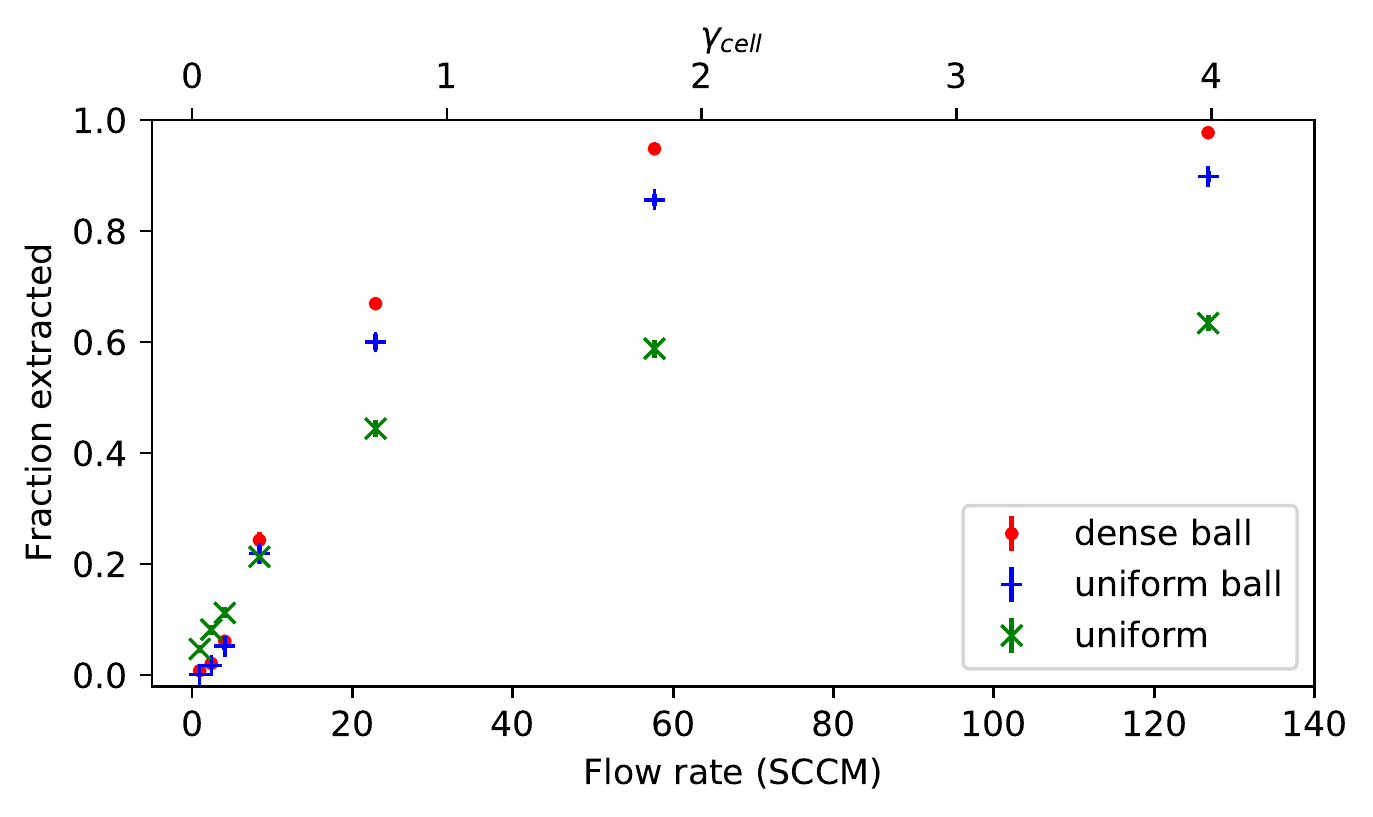}
\caption{Extraction versus 4 K He flow (SCCM) with three initial conditions: 1 cm diameter Gaussian ball, 1 cm diameter uniform ball, and uniform distribution over the cell. The initial molecules are thermalized at 4 K. The error bars are calculated using binomial statistics.}
\label{iniconcomparisonextractionrate}
\setlength{\belowcaptionskip}{-12pt}
\begin{center}
\includegraphics[width=0.47\textwidth]{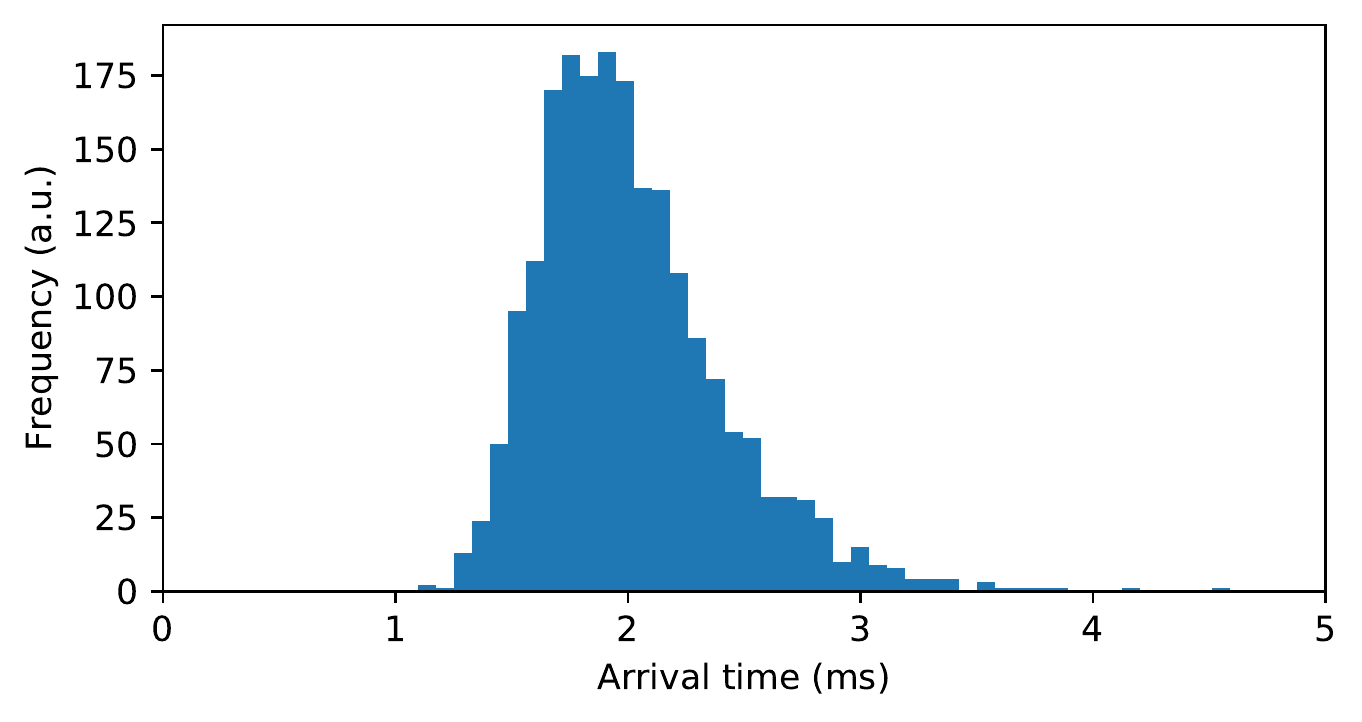}
\caption{Histogram of pumpout times for a 20 SCCM helium flow, or equivalently instantaneous number density at the aperture versus time after the molecules are created. The initial spatial distribution of molecules is a 1 cm diameter Gaussian ball.}
\label{arrivaltime}
\end{center}
\end{figure}

We studied the dependence of extraction on initial spatial distribution by comparing simulations with three different initial conditions: a 1 cm diameter Gaussian ball (with standard deviation 0.33 cm), a 1 cm diameter uniform ball, and a uniform distribution over the entire cell. The molecule velocities are initialized to a 4 K thermal distribution with zero mean velocity.  Figure \ref{iniconcomparisonextractionrate} shows the extraction fraction for each of these three cases under various flow rates. While the extraction fraction itself depends on these initial conditions, we see that the qualitative behavior is consistent among all of them. This behavior can can be understood by comparing the timescales of ``pumpout'' $\tau_{pump}$, the time it takes for a typical buffer gas atom to exit the cell, against diffusion $\tau_{dif\!f}$, the time it takes for a typical molecule to diffuse to the walls \cite{Patterson2007,Hutzler2012}. The ratio $\gamma_{cell} = \tau_{dif\!f}/\tau_{pump}$ therefore characterizes the extraction behavior.

For $\gamma_{cell} \lesssim 1$, most molecules will diffuse to the cell walls and stick. However, for $\gamma_{cell} \gtrsim 1$, many of the molecules will be extracted from the cell, and the extraction saturates to some maximum value for $\gamma_{cell}\gg 1$, as depicted in Fig. \ref{iniconcomparisonextractionrate}. Notice, however, that the maximum saturated values depend on the initial spatial distribution of molecules; this makes intuitive sense, as the fate of each molecule depends on its distance from a wall or the aperture.  This spatial dependence contributes to the typical molecular beam pulse width of $\sim1$ ms at the exit aperture, as shown in Fig. \ref{arrivaltime}, though it is possible to design and engineer cells to have shorter effective pumpout times to create narrower pulses \cite{Truppe2018}.

\begin{figure}[ht]
\setlength{\belowcaptionskip}{-12pt}
\begin{center}
\includegraphics[width=0.47\textwidth]{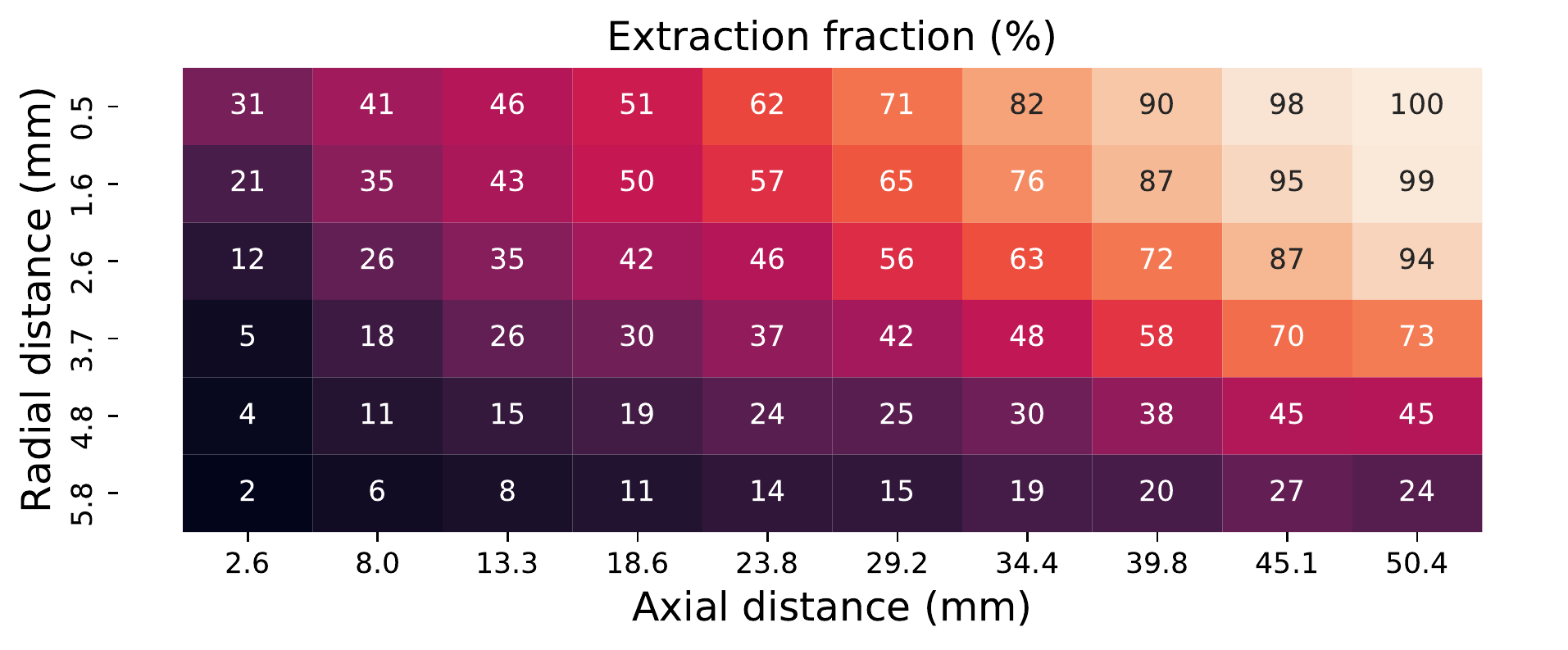}
\includegraphics[width=0.47\textwidth]{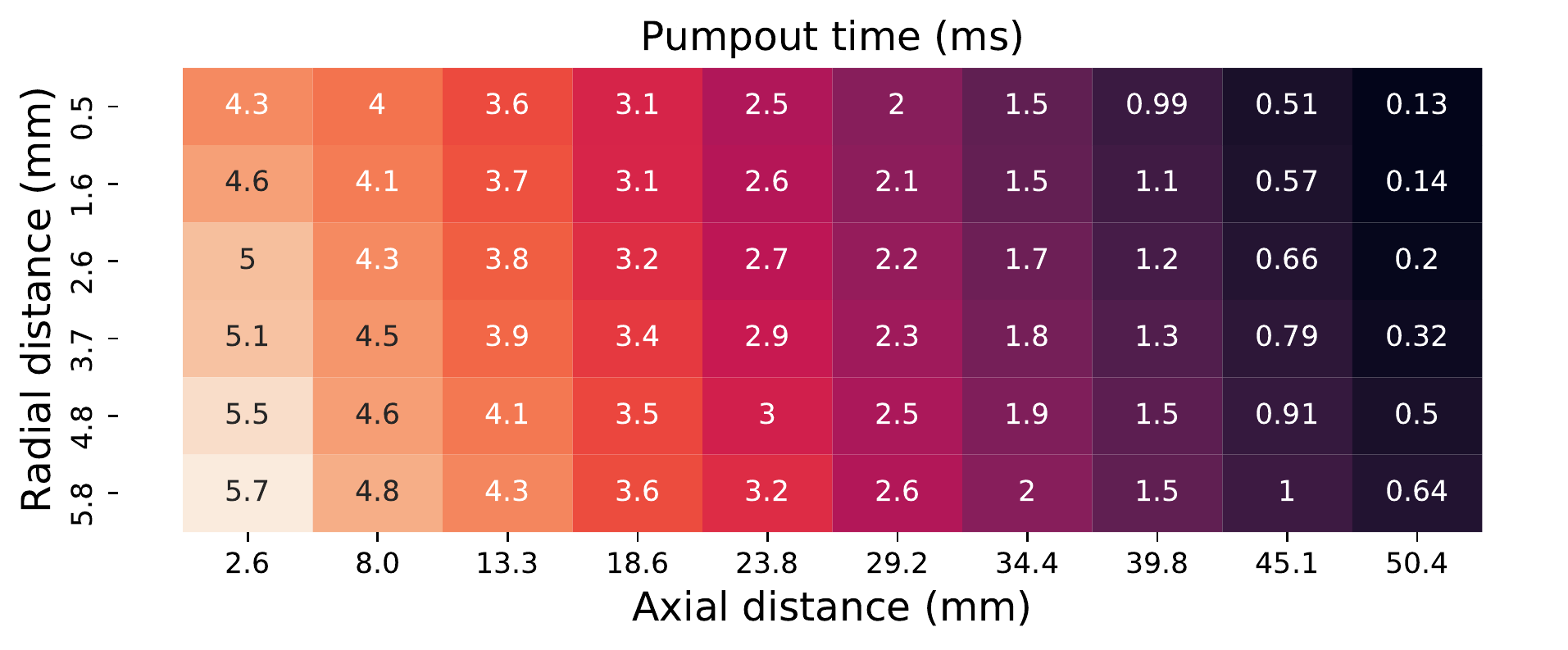}
\includegraphics[width=0.47\textwidth]{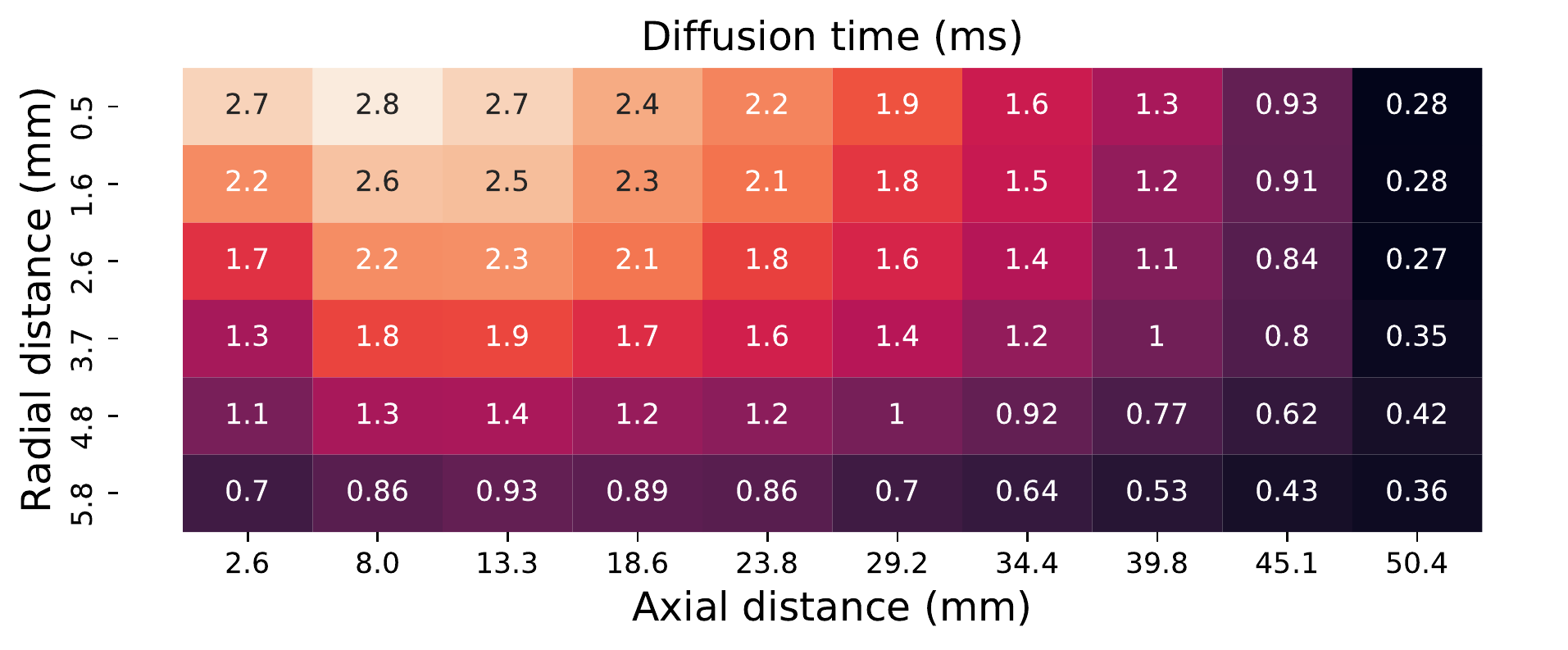}
\caption{Heat map of extraction fraction (top), pumpout time (middle), and diffusion time (bottom) versus initial molecule position for a 20 SCCM buffer gas flow. The $x$-axis indicates the distance from the point where the buffer gas inlet tube ends (0 mm) and finishes at the exit aperture (53 mm). The $y$-axis indicates the radial distance from the central axis of the cell. Therefore, the top right (left) corresponds to the exit aperture (fill line), and the bottom left and right are interior corners. Note that the $x$- and $y$- axes have different scales.}
\label{heatmap}
\end{center}
\end{figure}

To further elucidate this spatial dependence, we can map out the extraction probability, pumpout time, and diffusion time within the cell.  Figure \ref{heatmap} shows these quantities for a buffer gas flow rate of around 20 SCCM.  Note that the extraction drops fairly rapidly as one approaches the ``corner'' (lower right), possibly due to vortex formation, as has been noted elsewhere \cite{Truppe2018,Singh2018}. The diffusion time can be estimated \cite{Hutzler2012} to be around 3.3 ms, which is close to our results near the interface with the buffer gas inlet tube.

\subsubsection{Velocity}

\begin{figure}[ht]
\setlength{\belowcaptionskip}{-12pt}
\begin{center}
\includegraphics[width=0.47\textwidth]{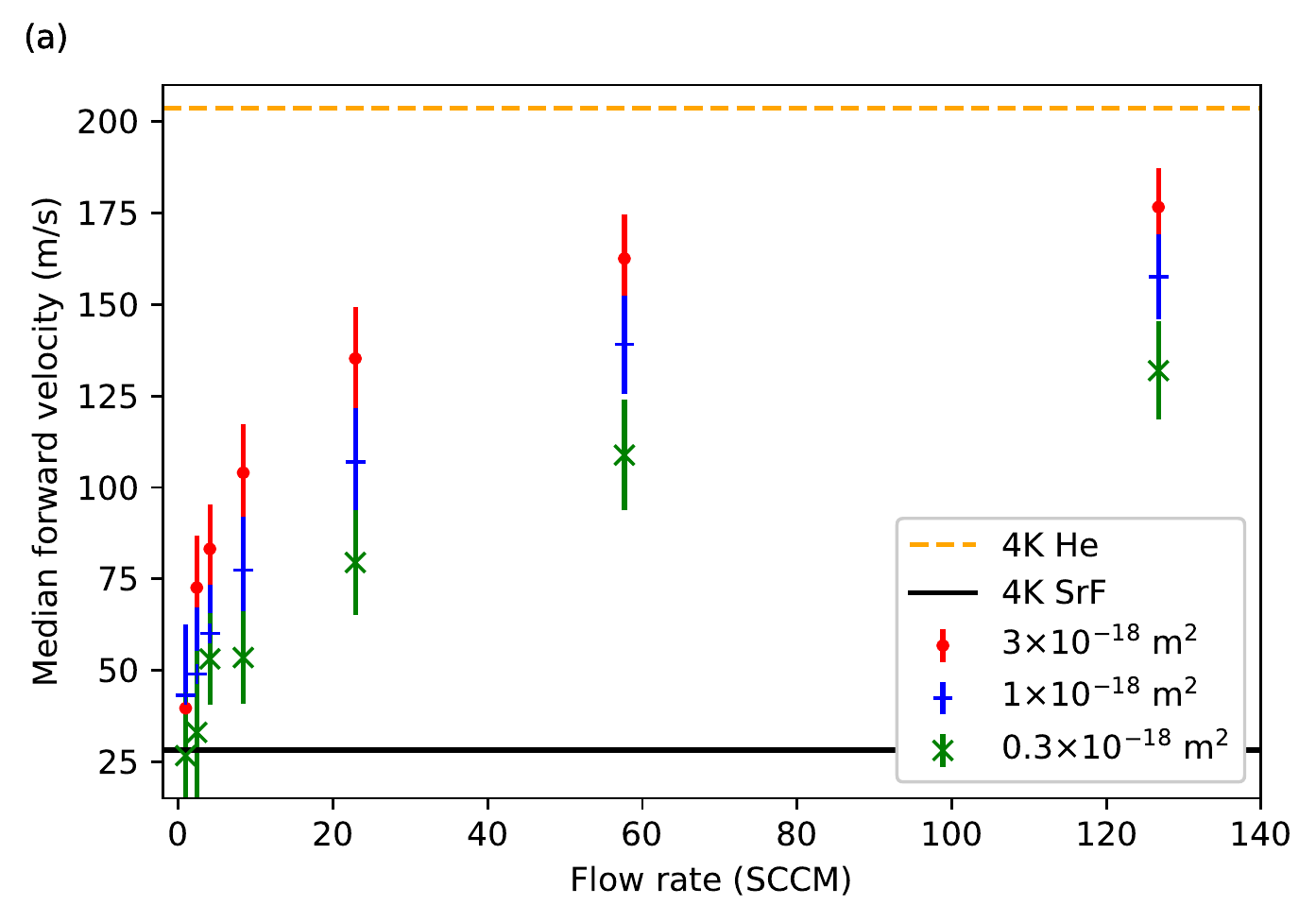}
\includegraphics[width=0.47\textwidth]{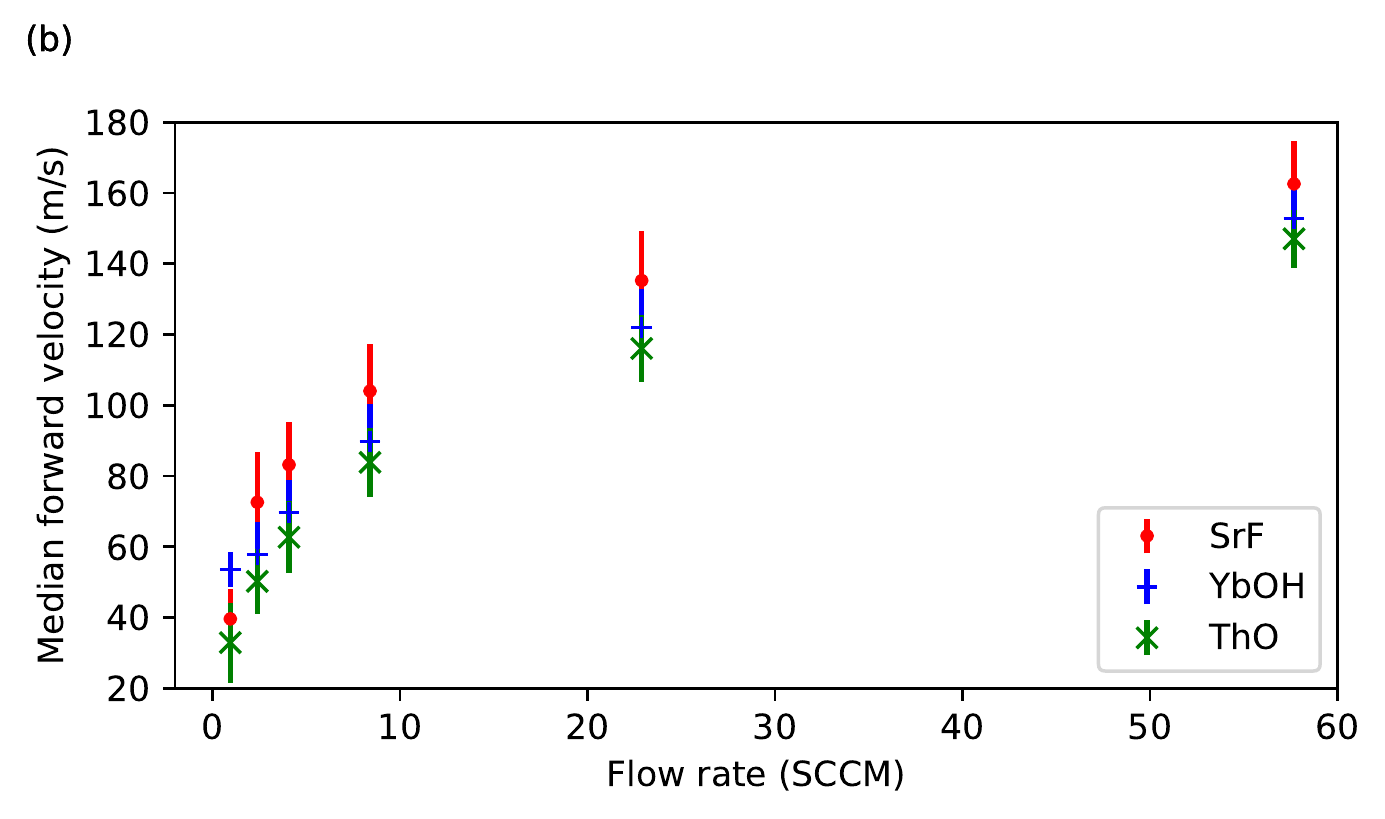}
\caption{(a) Median forward velocities versus 4 K He flow (SCCM) for SrF mass and various cross sections $\sigma_{b-s}$. The dashed and solid lines indicate 4 K He supersonic mean velocity and 4 K SrF thermal velocity, repectively. (b) Median forward velocities versus 4 K He flow (SCCM) for cross section $\sigma_{b-s}$ = $3\times 10^{-14}$ cm$^{2}$ and various molecular masses. The error bars indicate the standard deviation of the velocity distribution.}
\label{massmedianforward}
\end{center}
\end{figure}

Another critically important feature of the CBGB is the relatively low forward velocity of the molecular beam, typically $\lesssim$ 200 m/s.  This enables molecular beam experiments with long coherence times \cite{Baron2014,ACME2018} and efficient slowing of molecules to the capture velocity of magneto-optical traps \cite{Barry2012,Tarbutt2019,McCarron2018}.  Figure \ref{massmedianforward}(a) shows the median forward velocities versus buffer gas flow rate with different cross section values $\sigma_{b-s}$ = \{3, 1, 0.3\}$\times$10$^{-14} $ cm$^{2}$.  For all velocity plots, the reported values are measured 3 cm from the exit aperture. As the cross section value increases, the forward velocity also increases but its dependence on flow rate is the same, increasing linearly before saturating at some value. The behavior is consistent with results from experiments using a similar configuration \cite{Hutzler2011,Barry2011,Bulleid2013}, as well as theoretical expectations  \cite{Hutzler2012}.  For low flow, the source is effectively effusive and we expect a velocity close to that of an effuse thermal source of molecules.  At high flow, we expect the large collision rate to keep the molecules in equilibrium with the buffer gas, which approaches the supersonic velocity of $\approx200$ m/s for 4 K He. This increase of molecular forward velocity with increasing flow is typically called ``boosting.''

Our simulation shows lower velocity at low flows compared to experimental results for SrF \cite{Barry2011} and ThO \cite{Hutzler2011} but more closely matches those for Yb \cite{Bulleid2013}.  This discrepancy could be explained by a number of factors, including imperfect thermalization at low densities or instantaneous changes in densities as buffer gas is desorbed from the cell walls by ablation.  However, the higher flow behavior is consistent with the experimental data, particularly for the larger cross section value of $3\times 10^{-14}$ cm$^{2}$.  Note that the cross section could be treated as a fit parameter if the goal is to match data from a particular experiment.

\begin{figure}[ht]
\setlength{\belowcaptionskip}{-12pt}
\begin{center}
\includegraphics[width=0.47\textwidth]{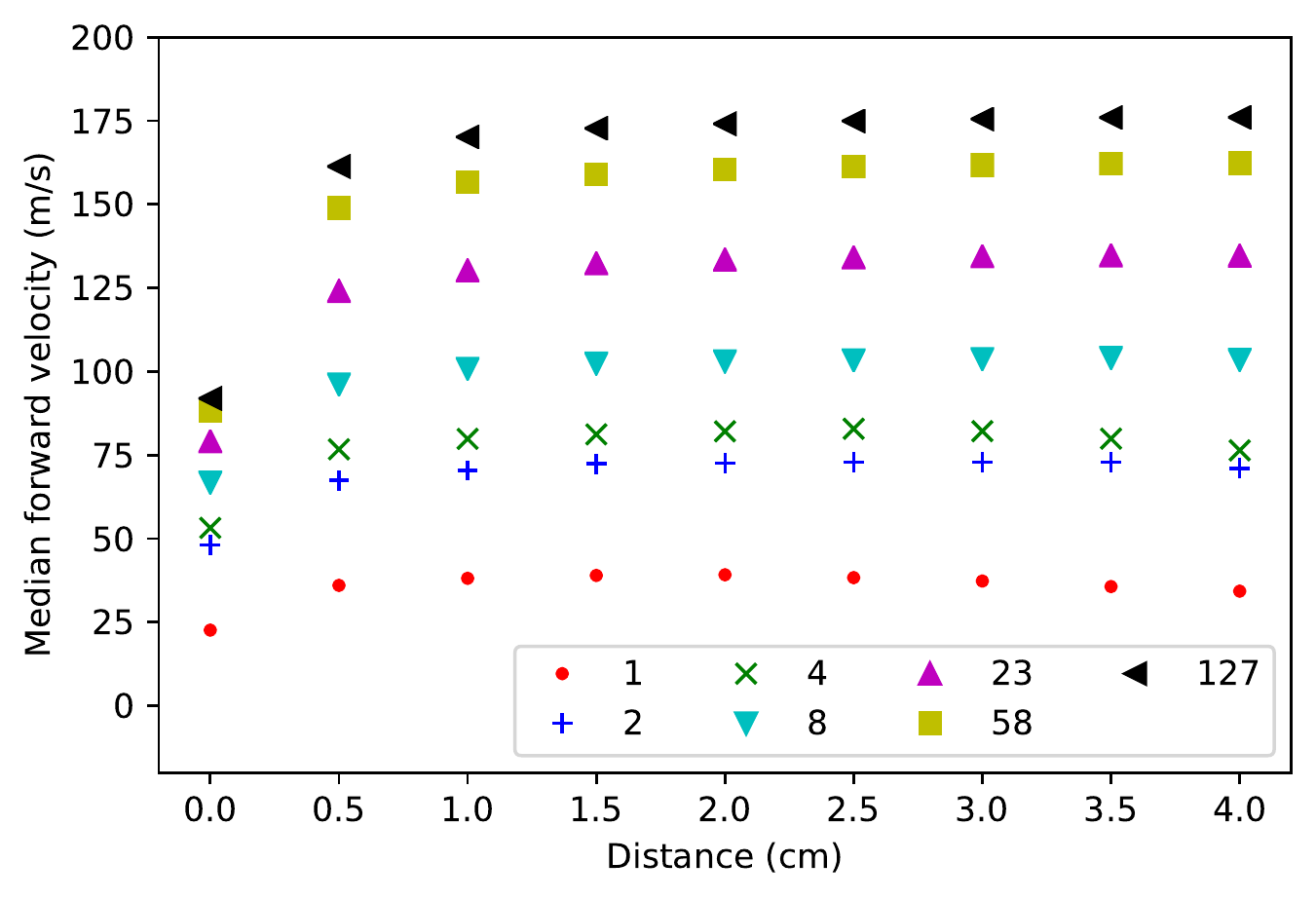}
\caption{Median forward velocities versus distance from the exit aperture for 4 K He with various flow rates (SCCM). The legend indicates flow rate (SCCM) rounded to the nearest integer.}
\label{forward_distance}
\end{center}
\end{figure}

Figure \ref{massmedianforward}(b) shows the median forward velocities versus He flow at 4 K for species of different mass: SrF, YbOH, and ThO. As the mass of the species decreases, the forward velocity increases while the general trend of velocity versus flow is maintained. However, the effect is not significant, which agrees with experimental data for species over this mass range \cite{Hutzler2011,Barry2011,Bulleid2013}.  Figure \ref{forward_distance} shows the median forward velocities versus distance from the exit aperture for 4 K He with various flow rates. The forward velocity saturates at around 1 cm from the exit aperture, similar to experimental data \cite{Barry2011}. 

We explored the sensitivity of these simulated velocities to various parameters of the buffer gas cell configuration, such as the thickness of the exit aperture (by a factor of 5, up to 2.5 mm) and the initial distribution of molecules, but these did not have an appreciable effect on the results. We also ran these simulations with 16 K He and 16 K Ne buffer gas and found that the behavior scaled as expected with the thermal velocities and temperature of the buffer gas atoms.

\subsubsection{Beam divergence}

\begin{figure}[ht]
\setlength{\belowcaptionskip}{-12pt}
\begin{center}
\includegraphics[width=0.47\textwidth]{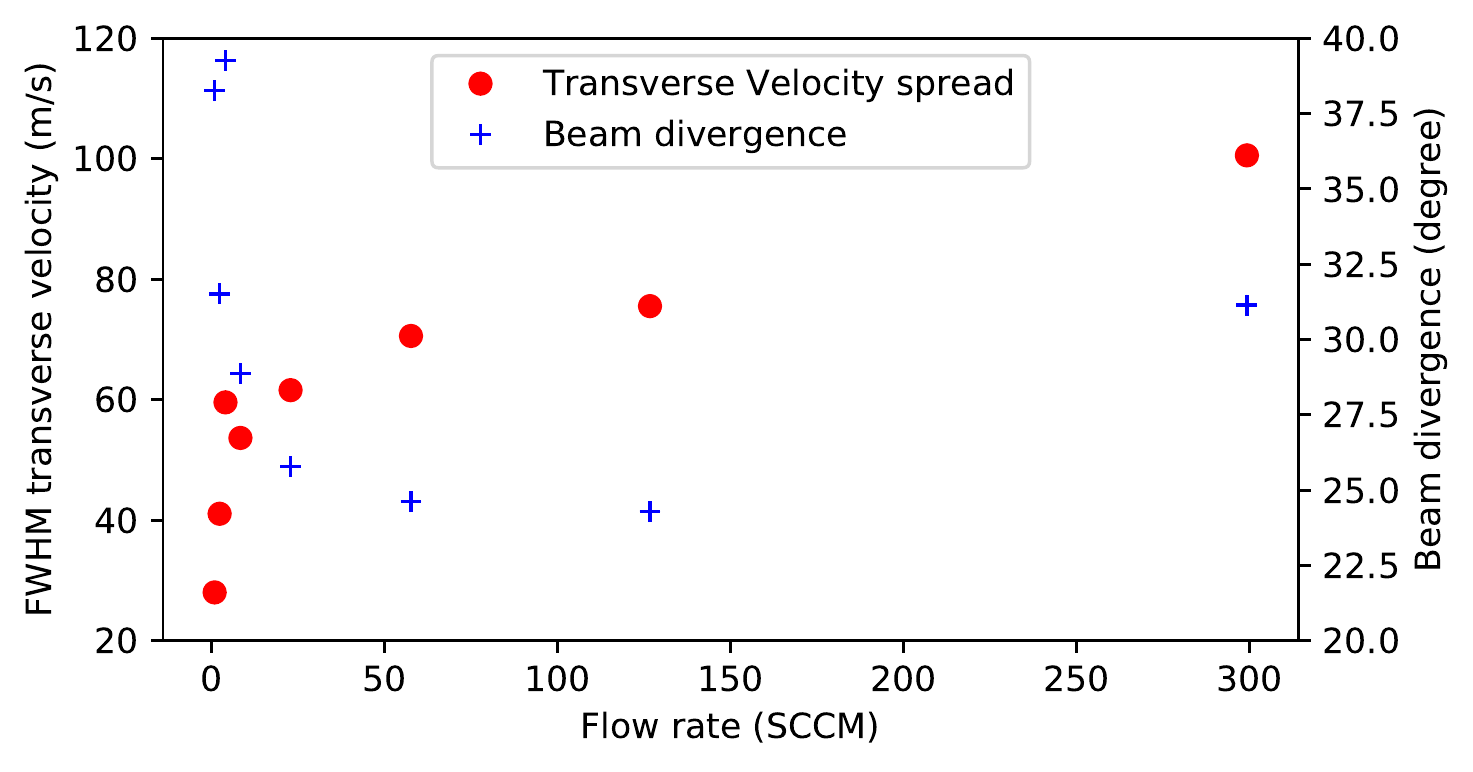}
\caption{Transverse velocity FWHM and beam divergence versus 4 K He flow (SCCM). The slope of the transverse velocity FWHM at higher flow rates is roughly 0.2 m/s/$\Re$, which is consistent with  \cite{Hutzler2011, Barry2011}.}
\label{singles}
\end{center}
\end{figure}

A third key feature of CBGBs is low beam divergence.  Molecules from a beam source are emitted into a finite solid angle, only a fraction of which can be used in an experiment, with the remainder leaving the beam line and striking a wall.  Lower divergence results in higher density on the beam line, and therefore higher useful molecular flux. A common measure of beam divergence is given by \cite{Hutzler2012}
\begin{equation}
	\Delta\theta = 2\,\text{arctan}\left(\frac{\Delta v_{\perp}}{2v_{\parallel}}\right),
\end{equation}
where $\Delta v_{\perp}$ is the transverse velocity spread (FWHM) and $v_{\parallel}$ is the mean forward velocity. Since the transverse spread is set by the molecular species' thermal velocity, while the forward velocity is largely set by the buffer gas thermal velocity, we expect the angular spread to be narrower than that of an effusive or supersonic source, where $\Delta v_{\perp}\sim v_{\parallel}$.  Figure \ref{singles} shows the beam divergence versus 4 K He flow for a single-stage cell. The minimum arises from the fact that the transverse velocity spread continues to increase while the forward velocity begins to saturate. The value at this minimum can be estimated \cite{Patterson2007,Hutzler2012,Bulleid2013} to be $\Delta\theta \sim 1-2\times\sqrt{m_b/m_s}$. For the case of SrF in He, this value ranges from $\approx 10-20^\circ$, compared to the simulated minimum of $\approx 25^\circ$.

Note, however, that the experimental data from different CBGB sources show differing behavior of divergence versus flow.  For example, ThO and SrF in He buffer gas \cite{Hutzler2011,Barry2011} show a flat divergence versus flow relationship and at a larger than expected value, ThO in Ne \cite{Hutzler2011} exhibits a minimum but at a larger than expected value, and Yb in He \cite{Bulleid2013} has a minimum close to the expected value. As we shall discuss in section \ref{delavalsec}, the kinetics of the expansion depend on the specific geometry of the nozzle \cite{Bulleid2013,Murphy1984}, which was similar but not identical in these various experimental implementations.

\subsection{Two-stage cell}

We now move on to discussing some variations of the traditional CBGB design, with the first being the two-stage ``slowing'' cell.  In these remaining sections, we will focus only on the salient features of these variants compared to the single-stage cell.

\begin{figure}[ht]
\setlength{\belowcaptionskip}{-12pt}
\begin{center}
\includegraphics[width=0.47\textwidth]{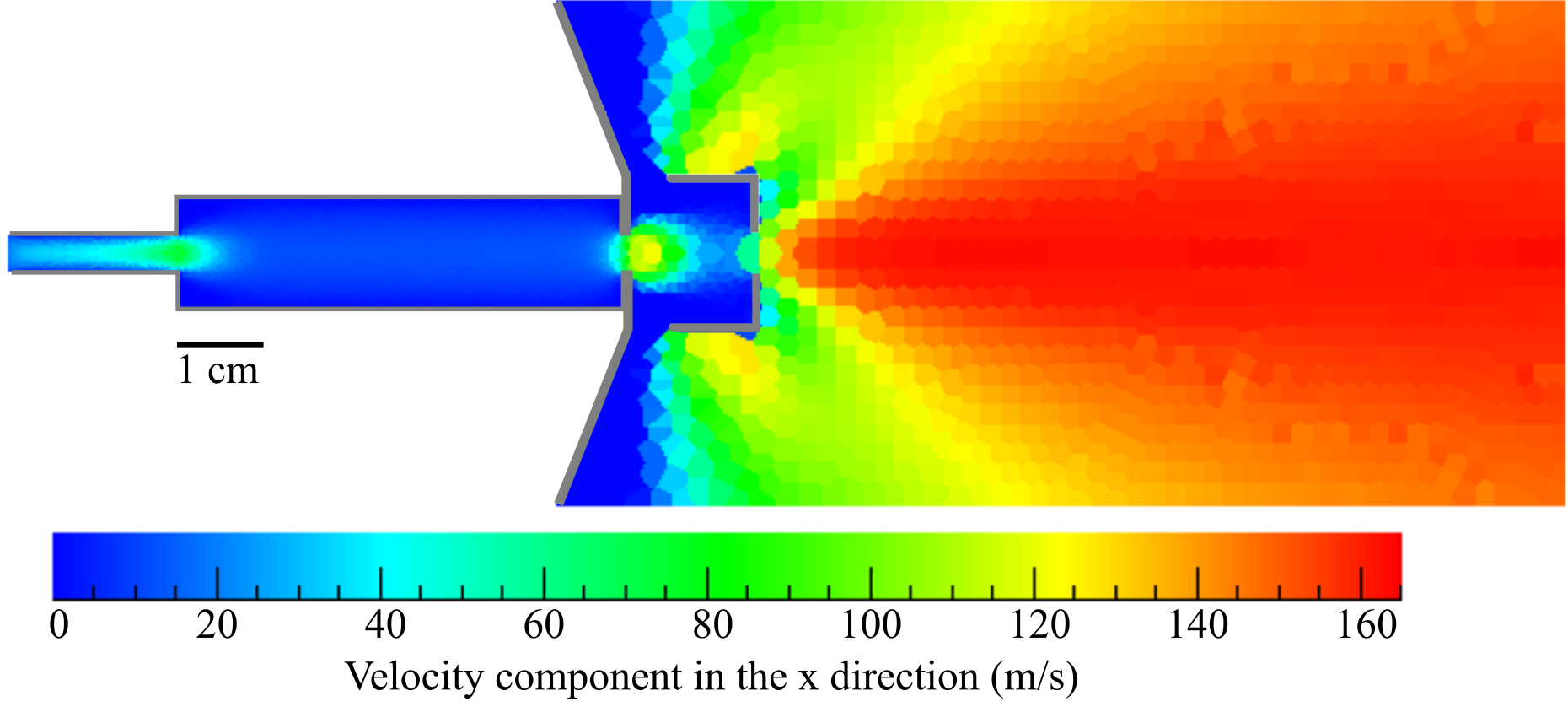}
\caption{Spatial variation of the forward velocity of the 4 K He buffer gas at around 20 SCCM in a two-stage cell as computed with DS2V. It consists of the same single-stage buffer gas cell as the previous section with the addition of a slowing cell.}
\label{pic2stage}
\end{center}
\end{figure}

The two-stage cell \cite{Patterson2007,Lu2011} is designed to benefit from hydrodynamic extraction while producing a slower molecular beam than a traditional single-stage cell.  The idea is to create a steady-state density of buffer gas at the aperture, thereby stifling the boosting effect through collisions with nearly stationary buffer gas atoms.  Our simulated cell consists of the same single-stage buffer gas cell as described in the previous section, with the addition of a second slowing cell after a 4.7 mm gap, as shown in Fig. \ref{pic2stage}.  This second cell has a 17 mm internal diameter and a length of 10 mm, with an exit aperture of 5 mm diameter and 0.5 mm thickness.  Note that experimental implementations of this design use a variety of configurations, several of which were examined in \cite{Lu2011}.

\begin{figure}[ht]
\setlength{\belowcaptionskip}{-12pt}
\begin{center}
\includegraphics[width=0.47\textwidth]{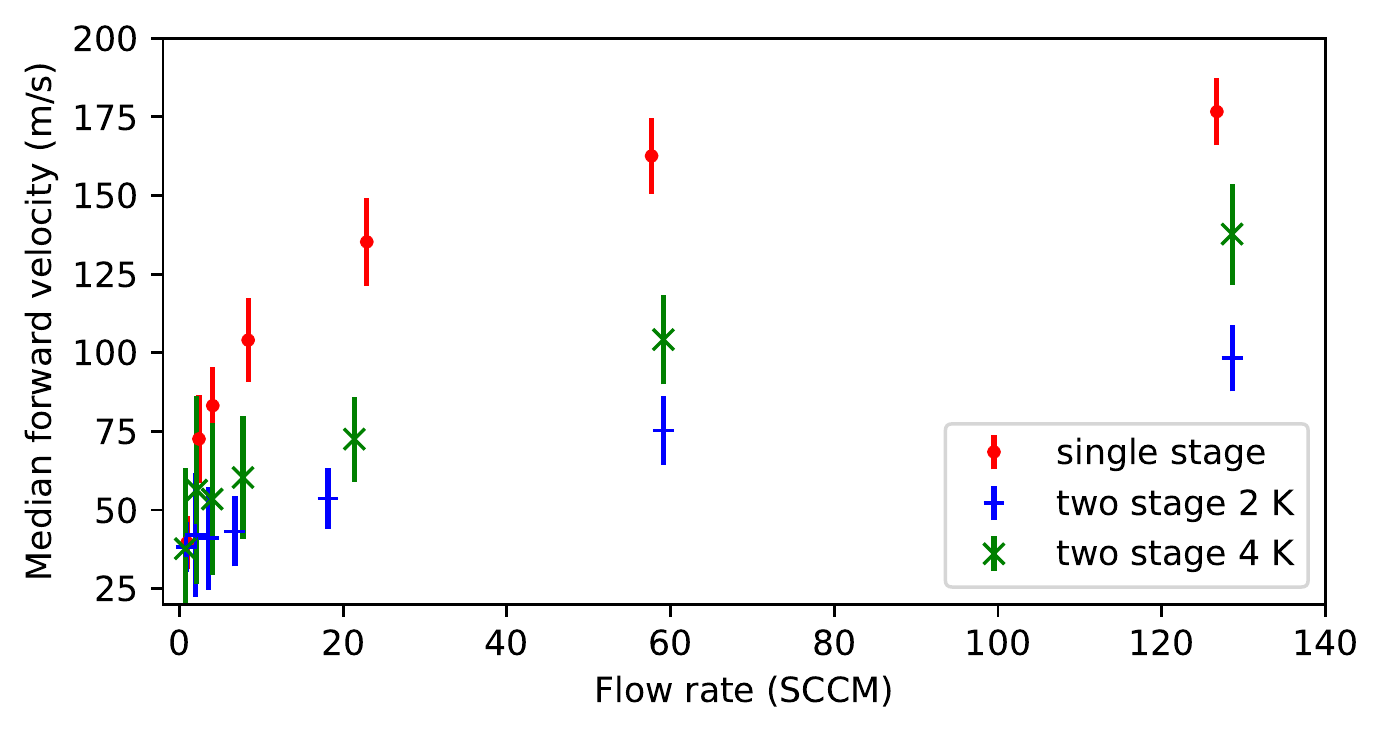}

\includegraphics[width=0.47\textwidth]{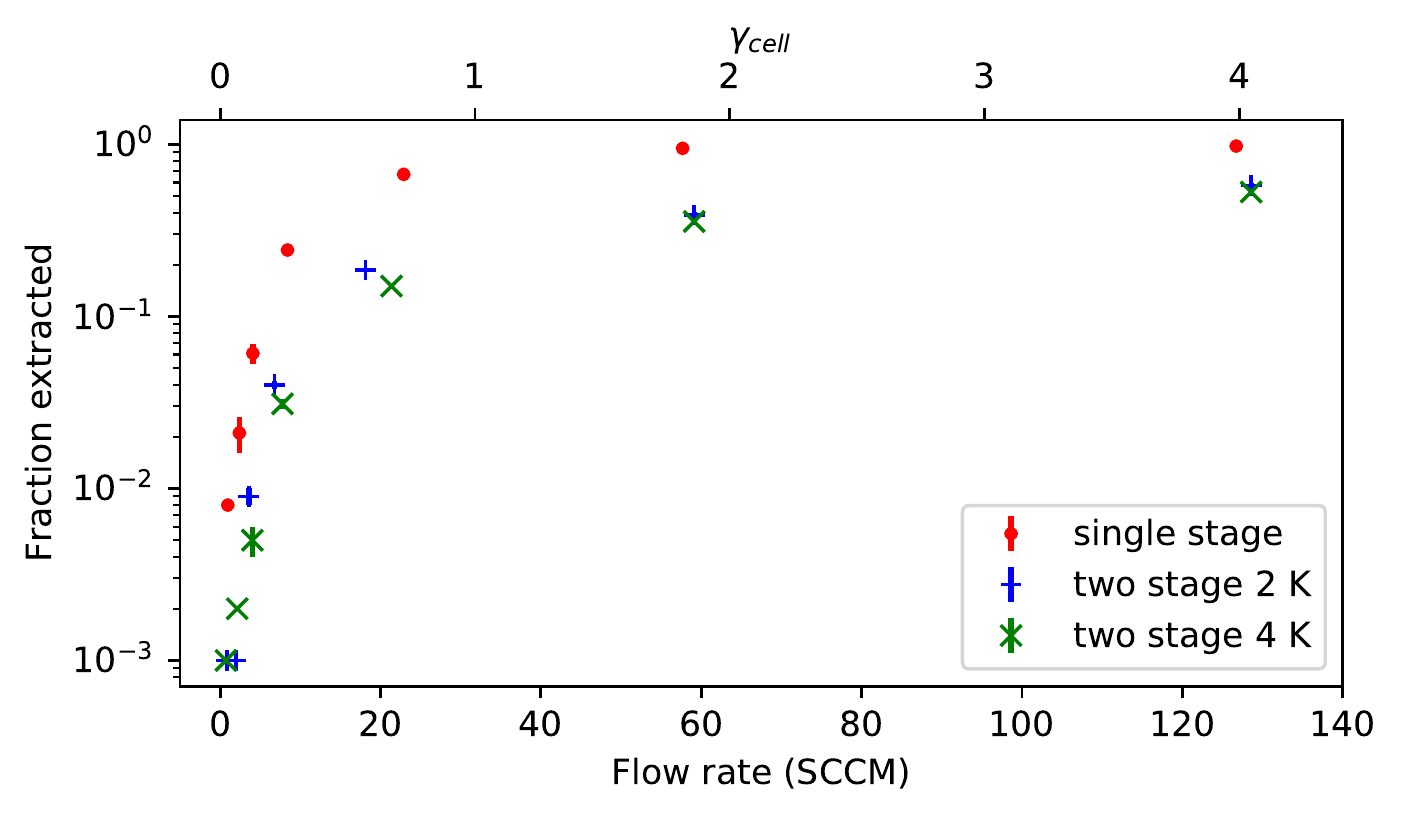}
\caption{Median forward velocities (top) and extraction fraction (bottom) versus He flow (SCCM) for single-stage (4 K) and two-stage (2 K and 4 K) cells.}
\label{2stages}
\end{center}
\end{figure}

Figure \ref{2stages} shows the median forward velocities and the extraction fractions versus 2 and 4 K He flow with a single-stage cell and two-stage cell. Compared to a single-stage design, the two-stage cell reduces the forward velocity by a factor of $\sim$1.5 but simultaneously reduces the extraction by up to a factor of $\sim10$. This is consistent with observations in experimental implementations of two-stage cells \cite{Lu2011}. Similarly, decreasing the temperature of the cell from 4 K to 2 K results in a reduction of the forward velocity by a factor of $\sim\sqrt{2}$, as expected. The boosted molecules at the first exit aperture experience several collisions with nearly-stationary buffer gas atoms and are slowed.  The density of the buffer gas is also much lower in the second stage than the first stage (roughly by a factor of 10), so the boosting effect at the second aperture is weaker.  Thus, the molecules exit the second aperture slower than in a traditional single stage cell.

\subsection{\label{delavalsec}de Laval nozzle}

\begin{figure}[ht]
\setlength{\belowcaptionskip}{-12pt}
\begin{center}
\includegraphics[width=0.47\textwidth]{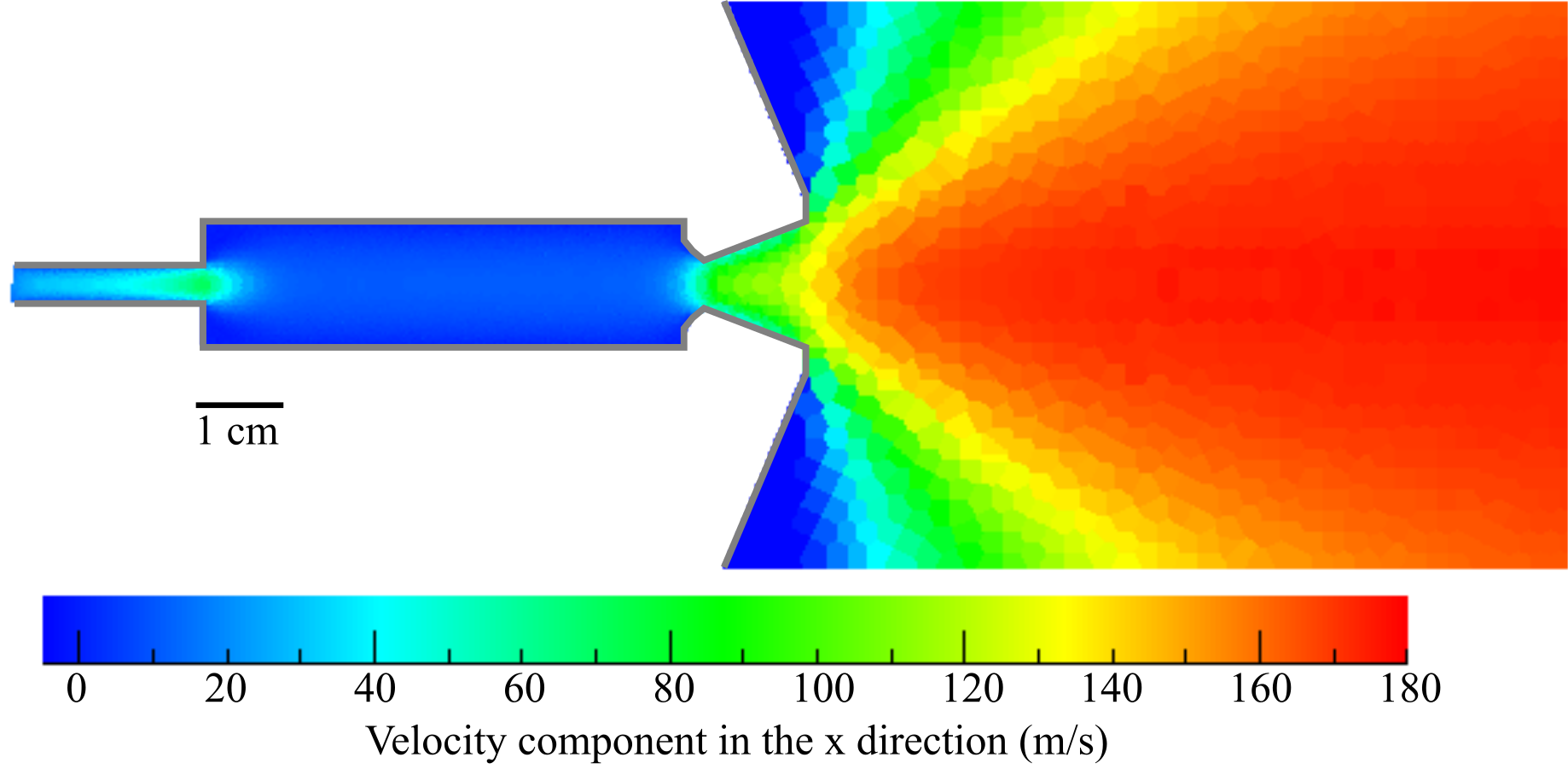}
\caption{Spatial variation of the forward velocity of the 4 K He buffer gas at around 10 SCCM in a cell with a de Laval nozzle, as computed with DS2V.}
\label{picdelaval}
\end{center}
\end{figure}

\begin{figure}[ht]
\setlength{\belowcaptionskip}{-12pt}
\begin{center}
\includegraphics[width=0.47\textwidth]{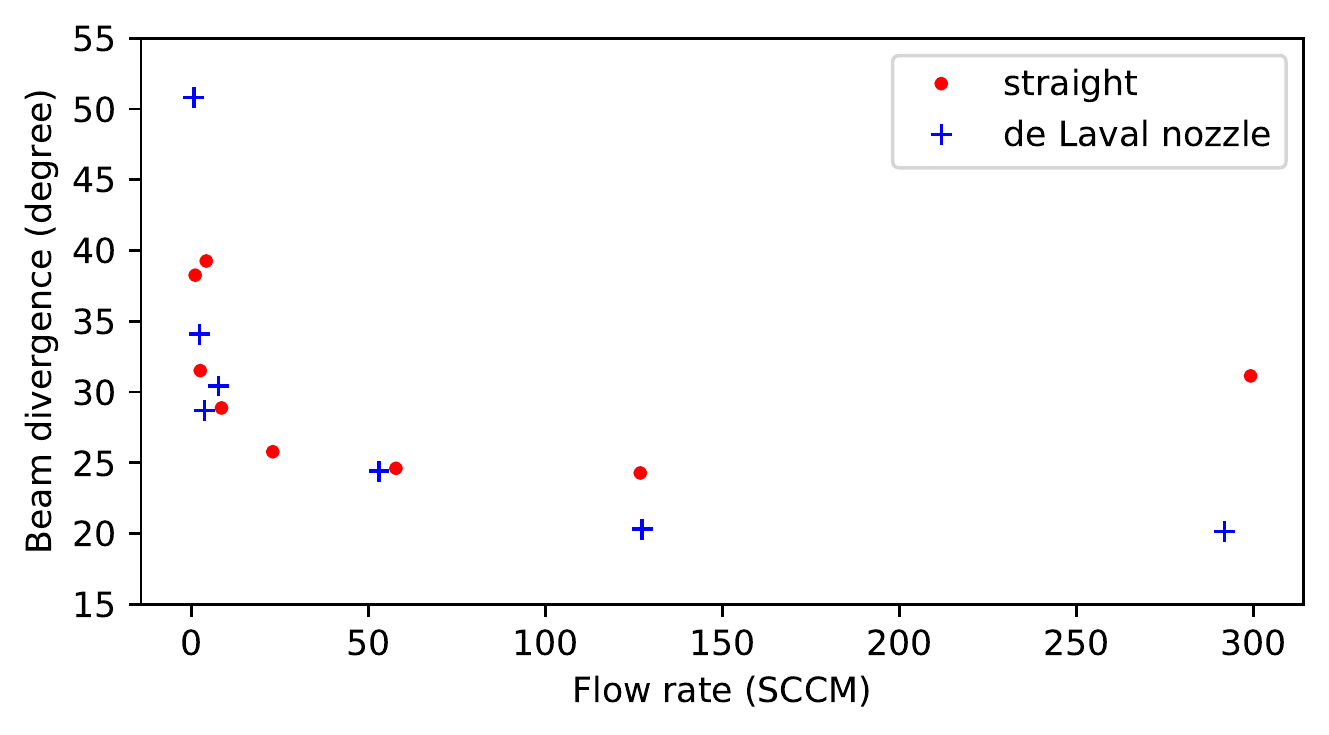}
\includegraphics[width=0.47\textwidth]{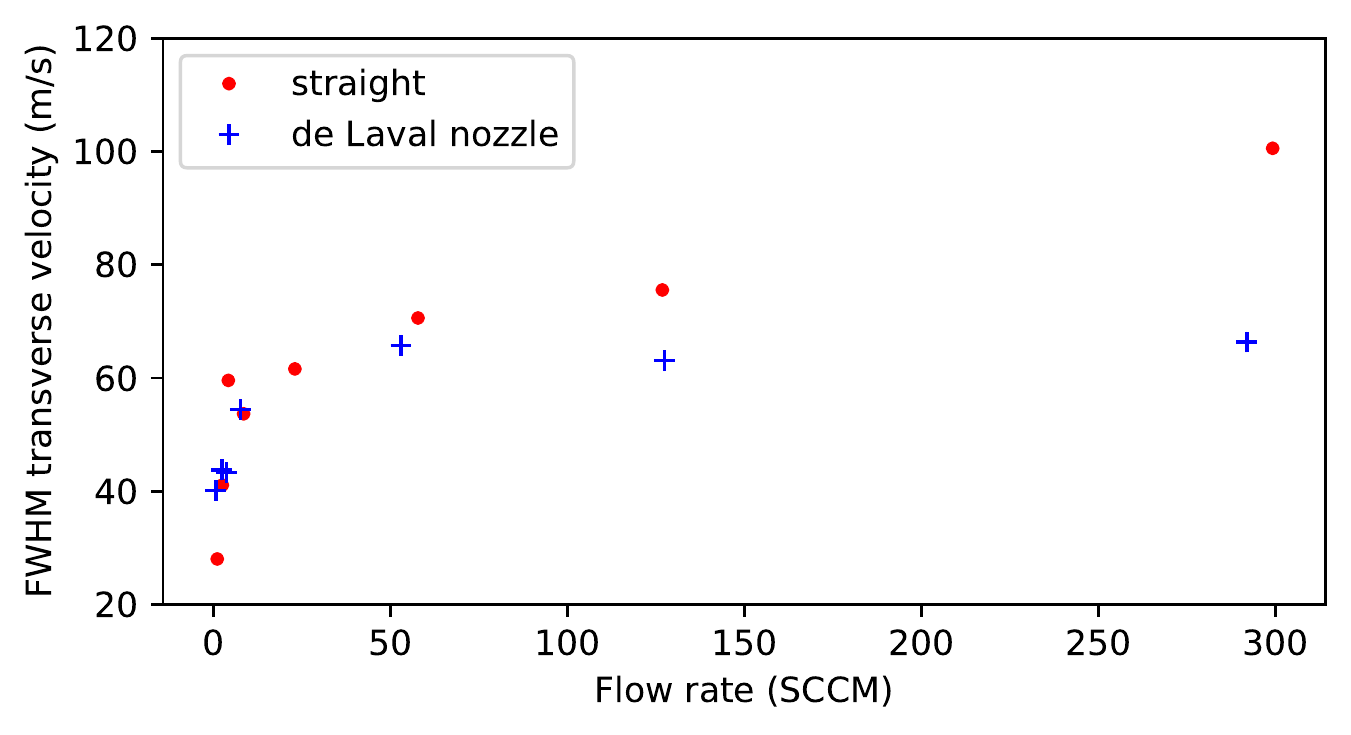}
\caption{Beam divergence (top) and transverse velocity spread (bottom) versus 4 K He flow (SCCM) for the straight (standard single-stage) and de Laval apertures.}
\label{delavals}
\end{center}
\end{figure}

A de Laval nozzle \cite{Pauly2000,Sutton2016,Xiao2018} is a converging-diverging nozzle designed to reduce the transverse velocity spread and beam divergence in high flow regimes without significantly reducing the overall extraction fraction. The de Laval cell is a modified single-stage buffer gas cell with a de Laval nozzle placed at the exit aperture \cite{Xiao2018}, as shown in Fig. \ref{picdelaval}. We use the geometry of \cite{Xiao2018} but approximate the shape of the nozzle with four straight lines.  Figure \ref{delavals} compares the beam divergence and transverse velocity spread for a traditional single-stage cell versus a cell with a de Laval aperture. The divergence of the molecular beam is reduced compared to the standard aperture at the high flow rates ($\gtrsim $ 50 SCCM), where the hydrodynamic effects of the de Laval nozzle are expected to be important, as seen in previous experimental results \cite{Xiao2018}. The reduction in divergence results almost entirely from the lower transverse velocity spread, as we find both in the simulations and the experimental data that the forward velocity is essentially unaffected.

\section{Outlook}

We have implemented a two-step approach to simulating cryogenic buffer-gas beams and used it to demonstrate non-trivial features in a number of different geometries. While this manuscript has focused on reproducing and understanding known behavior, the real utility of our work will be to numerically optimize geometries for a given purpose.  Fluid dynamics simulations have already been used to successfully design cells with enhanced properties such as extraction \cite{Truppe2018,Singh2018}, and we hope that the current approach will be especially useful for beam properties that depend critically on both high and low density regimes, such as beam divergence and forward velocity.  Furthermore, the output particles can be fed into simulations of post-CBGB techniques, such as laser slowing and cooling \cite{Tarbutt2015PRA}, Stark \cite{Bethlem1999} and Zeeman \cite{Vanhaecke2007,Narevicius2008} deceleration, guiding \cite{Patterson2007,VanBuuren2009}, and beam loading \cite{Egorov2002,Sprouse1989}.  The ability to numerically fine-tune buffer-gas cells for specific applications will enable rapid prototyping of CBGB implementations and open new directions in molecular quantum control.

\begin{acknowledgments}

We appreciate many helpful discussions with the PolyEDM collaboration, especially Ben Augenbraun and Cal Miller.  We are grateful for feedback on the manuscript from Arian Jadbabaie and Phelan Yu. Y. T. was supported by the Masason Foundation.  D. S. was supported by a Caltech Summer Undergraduate Research Fellowship (SURF) sponsored by the Aerospace Corporation and Gary Stupian.  G. W. was supported by SURF and the Heising-Simons Foundation (2019-1193).  N. R. H. acknowledges support from an NSF CAREER award (PHY-1847550), a NIST Precision Measurement Grant (60NANB18D253), the Gordon and Betty Moore Foundation (7947), and the Alfred P. Sloan Foundation (G-2019-12502).  Computations in this manuscript were performed on the Caltech High Performance Cluster.

\end{acknowledgments}

\appendix

\section{\label{bayes_sec}Bayesian Thermal Velocity Sampling}

We will work in the rest frame of a molecule moving through a buffer gas.  The thermal velocities of buffer gas atoms \textbf{u} = (u, $\theta$, $\varphi$) in spherical coordinates are distributed in accordance with the Maxwell-Boltzmann distribution (Eq. \ref{boltzmann}).
However, these atoms are not all equally likely to collide with a molecule, as the collision rate between particles scales with their relative velocity. To capture this effect when sampling a buffer gas velocity at the time of a collision, we employ the Bayes-updated distribution,
\begin{equation}
\begin{split} \label{bayes}
    f(\textbf{u}|\text{coll}) &\propto P(\text{coll}|\textbf{u}) \times f_\text{MB}(\textbf{u}) \\
    &\propto |\textbf{v}_\text{flow} + \textbf{u}| \times u^2 e^{-\frac{mu^2}{2kT}}\sin\theta,
\end{split}
\end{equation}
where we have used $f_\text{MB}(\theta) \propto \sin\theta$ and defined $\textbf{v}_\text{flow}$ as the mean flow velocity of the surrounding buffer gas. Since we have chosen to work in the molecule's rest frame, the sum of thermal and flow velocities in the first factor of Eq. (\ref{bayes}) represents the relative velocity between the molecular species and a buffer gas atom. Note that in this frame, $\textbf{v}_\text{flow}$ is offset from its lab-frame value referenced in Eq. (\ref{vflow_lab}) of the main text and Appendix \ref{boyd}.

We will work in a coordinate system where $\hat{\textbf{z}}$ is aligned with $\textbf{v}_\text{flow}$, so that this factor can be expressed in terms of the coordinates of $\textbf{u}$ as
\begin{align}
    |\textbf{v}_\text{flow} + \textbf{u}| &= \sqrt{u^2 + 2uv_\text{flow}\cos\theta + (v_\text{flow})^2}.
\end{align}
As a result, we have that
\begin{align} \label{beysian}
    f(\textbf{u}|\text{coll}) &\propto u^3 e^{-\frac{mu^2}{2kT}}\sin\theta\sqrt{1 + 2\frac{v_\text{flow}}{u}\cos\theta + \left(\frac{v_\text{flow}}{u}\right)^2}.
\end{align}
We see that in any flow regime, the velocity distribution of atoms involved in collisions will be biased by an extra factor of $u$ when compared to the Maxwell-Boltzmann distribution \cite{Boyd2017}, and when there is a velocity slip $v_\text{flow}$ that is not negligible compared to $u$ (\textit{e.g.} near or beyond the aperture of a buffer gas cell), there are further adjustments contained in the square root factor that bias both the speeds and directions of colliding atoms. 

In our code, we sample from this velocity distribution by first generating a speed $u^*$ from the marginal distribution of $u$,
\begin{multline}
    f(u|\text{coll}) \propto \\ \int_0^\pi f(\textbf{u}|\text{coll}) d\theta \propto u e^{-\frac{mu^2}{2kT}} [(u + v_\text{flow})^3 - |u - v_\text{flow}|^3].
\end{multline}
Then, we can sample a polar angle $\theta^*$ from
\begin{multline}
    f(\theta|u^*, \text{coll}) \propto \\ (u^*)^3 e^{-\frac{m(u^*)^2}{2kT}}\sin\theta \sqrt{1 + 2\frac{v_\text{flow}}{u^*}\cos\theta + \left(\frac{v_\text{flow}}{u^*}\right)^2}.
\end{multline}

Finally, we can extract the components of our sample thermal velocity \textbf{u} $= (u^*, \theta^*, \varphi\in \text{Unif[0, 2}\pi])$ along the physical coordinate axes of our simulation $\{x', y', z'\}$ via the following rotation matrix: 
\begin{equation} \label{matrix_vel}
\begin{pmatrix}
    u_{x'} \\
    u_{y'} \\
    u_{z'}
\end{pmatrix}
=
\begin{pmatrix}
    \frac{\hat{v}_{y'}^2 + \hat{v}_{x'}^2\hat{v}_{z'}}{\hat{v}_{x'}^2 + \hat{v}_{y'}^2} & \frac{\hat{v}_{x'} \hat{v}_{y'} (\hat{v}_{z'} - 1)}{\hat{v}_{x'}^2 + \hat{v}_{y'}^2} & \hat{v}_{x'} \\
    \frac{\hat{v}_{x'} \hat{v}_{y'} (\hat{v}_{z'} - 1)}{\hat{v}_{x'}^2 + \hat{v}_{y'}^2} & \frac{\hat{v}_{x'}^2 + \hat{v}_{y'}^2\hat{v}_{z'}}{\hat{v}_{x'}^2 + \hat{v}_{y'}^2} &  \hat{v}_{y'} \\
    -\hat{v}_{x'} & -\hat{v}_{y'} & \hat{v}_{z'}
\end{pmatrix}
\begin{pmatrix}
    u_x \\
    u_y \\
    u_z
\end{pmatrix}
.
\end{equation}
Here, the sample velocity \textbf{u} has been expressed in Cartesian coordinates, and $\hat{v}_{i'} \equiv \frac{\textbf{v}_\text{flow}}{v_\text{flow}} \cdot \hat{\textbf{i}}'$ are the projections of the flow velocity unit vector (which defined the direction $\hat{\textbf{z}}$ in the unprimed coordinate system) onto the physical coordinate axes.
 This transformation corresponds to a passive rotation mapping $\hat{\textbf{z}} \mapsto \hat{\textbf{z}}^{\prime}$ about the axis $\hat{\textbf{n}} \propto \textbf{z} \times \hat{\textbf{z}}^{\prime}$.

\-\
\section{\label{vrel_sec}Average Relative Velocity}

Continuing to work in the rest frame of the molecular species and using the same definitions as in the previous section, we can write the average relative velocity between the molecule and surrounding buffer gas atoms as $\langle v_\text{rel} \rangle = \langle |\textbf{v}_\text{flow} + \textbf{u}| \rangle.$ We can average over the probability distribution of $\textbf{u}$ to evaluate the expectation value of $v_\text{rel}$ as follows:

\vspace{7mm}

\begin{widetext}
\begin{equation} \label{vrel}
\begin{split}
    \langle v_\text{rel} \rangle &= \left\langle \sqrt{u^2 + 2u v_\text{flow}\cos\theta  + (v_\text{flow})^2} \-\ \right\rangle \\
    &= \int_0^\infty du f(u) \int_{-1}^{1} d(\cos\theta) f(\cos\theta) \int_0^{2\pi}d\varphi f(\varphi) \sqrt{u^2 + 2u v_\text{flow}\cos\theta  + (v_\text{flow})^2} \\
    &= \sqrt{\frac{2kT}{m\pi}} \exp\left[-\frac{m(v_\text{flow})^2}{2kT}\right] + \left(\frac{kT}{mv_\text{flow}} + v_\text{flow}\right)\text{Erf}\left(v_\text{flow}\sqrt{\frac{m}{2kT}}\right).
\end{split}
\end{equation}
\end{widetext}
To evaluate the integrals in the second line, we took the thermal velocities $\textbf{u}$ to obey the Maxwell-Boltzmann distribution, which can be parameterized as

\begin{equation} \label{boltzmann}
\begin{split}
	f_\text{MB}(u) &= \left(\frac{m}{2\pi kT}\right)^{3/2} 4\pi u^2 e^{-\frac{mu^2}{2kT}} \\
	f_\text{MB}(\cos\theta) &= 1/2 \\
	f_\text{MB}(\varphi) &= 1/(2\pi),
\end{split}
\end{equation}
where $m$ and $T$ are the mass and temperature of surrounding buffer gas atoms and $k$ is the Boltzmann constant.

Within the buffer gas cell, frequent collisions keep the molecular species moving in approximate equilibrium with the buffer gas flow, so $v_\text{flow}$ is effectively a measure of fluctuations arising from the molecule's thermal velocity. Because the molecule is heavier than the buffer gas, we have that $m(v_\text{flow})^2 \ll kT$ in this regime, and the expression in Eq. (\ref{vrel}) simplifies to the mean Maxwell-Boltzmann velocity,
\begin{equation}
	\langle u \rangle = 2\sqrt{\frac{2kT}{m\pi}}.
\end{equation}

Near and beyond the aperture, however, the velocity slip between the molecular species and the buffer gas may be amplified and contribute significantly to $v_\text{flow}$. In this regime, therefore, it is especially important to consider both the thermal and flow contributions to the average relative velocity, as derived in Eq. (\ref{vrel}).

\section{Hard-Sphere Collision Mechanics} \label{boyd}
Once the thermal velocity \textbf{u} of a colliding buffer gas atom has been sampled from the distribution (\ref{beysian}) and converted to physical coordinates using Eq. (\ref{matrix_vel}), we can compute the impulse it imparts on the molecule of interest. 
This computation is done in the lab frame, where the molecular species has pre-collision velocity $\textbf{v}_s$ and the buffer gas atom has pre-collision velocity $\textbf{v}_b = \textbf{u} + \textbf{v}_\text{flow}$ (the flow velocity here refers to its value in the lab frame).
The center-of-mass velocity is therefore
\begin{equation}
	\textbf{v}_{CM} = \frac{m_s}{m_s+m_b}\textbf{v}_s + \frac{m_b}{m_s+m_b}\textbf{v}_b.
\end{equation}

Assuming a uniformly distributed impact parameter allows us to generate random scattering angles as \cite{Boyd2017}
\begin{equation}
\begin{split}
	\cos\theta &= 2R_1 - 1 \\
	\sin\theta &= \sqrt{1 - \cos^2\theta} \\
	\varphi &= 2\pi R_2,
\end{split}
\end{equation}
where $R_1$ and $R_2$ are independent random variables uniform on [0, 1]. These scattering angles determine the post-collision relative velocity components,
\begin{equation}
\begin{split}
	g_x &= g \sin\theta \cos\varphi	 \\
	g_y &= g \sin\theta \sin\varphi \\
	g_z &= g \cos\theta,
\end{split}
\end{equation}
where the magnitude of the relative velocity $g$ is unchanged from its pre-collision value of $|\textbf{v}_s - \textbf{v}_b|$. Finally, the post-collision velocity of the molecular species can be written as 
\begin{equation}
	\textbf{v}_s' = \textbf{v}_{CM} + \frac{m_b}{m_s + m_b}\textbf{g}.
\end{equation}

\bibliography{biblio}

\end{document}